\documentclass[aps,prb,twocolumn,showpacs,superscriptaddress]{revtex4}
\usepackage{graphicx}
\usepackage{color}
\renewcommand{\vec}[1]{\mathbf{#1}}
\begin{document}
\title{Complex phases in the doped two-species bosonic Hubbard Model}

\author{Kalani Hettiarachchilage} 
\affiliation{Department of Physics and Astronomy, Louisiana State University, Baton Rouge, Louisiana 70803, USA}
\affiliation{Center for Computation and Technology, Louisiana State University, Baton Rouge, LA 70803, USA}
\author{Val\'ery G.~Rousseau}
\affiliation{Department of Physics and Astronomy, Louisiana State University, Baton Rouge, Louisiana 70803, USA}
\author{Ka-Ming Tam}
\affiliation{Department of Physics and Astronomy, Louisiana State University, Baton Rouge, Louisiana 70803, USA}
\affiliation{Center for Computation and Technology, Louisiana State University, Baton Rouge, LA 70803, USA}
\author{Mark Jarrell}
\affiliation{Department of Physics and Astronomy, Louisiana State University, Baton Rouge, Louisiana 70803, USA}
\affiliation{Center for Computation and Technology, Louisiana State University, Baton Rouge, LA 70803, USA}
\author{Juana Moreno}
\affiliation{Department of Physics and Astronomy, Louisiana State University, Baton Rouge, Louisiana 70803, USA}
\affiliation{Center for Computation and Technology, Louisiana State University, Baton Rouge, LA 70803, USA}

\date{\today}

\begin{abstract}
We study a doped two-dimensional bosonic Hubbard model with two hard-core species using quantum Monte 
Carlo simulations. With doping we find five distinct phases, including a normal liquid at higher 
temperature, an anti-ferromagnetically ordered Mott insulator, a region of coexistent 
anti-ferromagnetic and superfluid phases near half filling, and further away from half filling, a superfluid 
phase and a phase separated ferromagnet. In the latter, the heavy species has Mott behavior with integer 
fillings, while the light species shows Mott and superfluid behaviors. The global entropy of this phase is
relatively high which may provide a new avenue to obtain a polarized phase or a Mott insulator in cold atom 
experiments.
\end{abstract}

\maketitle
Cold atoms experiments~\cite{Greiner} have become a playground for realizations of the 
Hubbard~\cite{Fisher, Batrouni} and other strongly correlated model Hamiltonians, since 
model parameters can be tuned using laser and magnetic fields.~\cite{Timmermans99, kohler:1311}
Recently, there has been an increasing interest in studies of mixtures of atoms~\cite{Modugno, Roati} 
due to the complexity associated with multiple species and the possibility of discovering novel 
phases. The experimental study of the $^{87}Rb$-$^{41}K$, $^{6}Li$-$^{40}K$ and different alkaline 
earth mixtures in an optical lattice ~\cite{Catani, Taie,Taglieber}  have motivated theoretical 
studies of the Hubbard model with two species with different masses.~\cite{Altman, Soyler, Stephen, Andrii, Rigol} 
These studies reveal a rich phase diagram 
at half-filling. Experimental studies of the two species Bose-Hubbard model shows that the massive 
species exhibits Mott while the other shows superfluid behavior.~\cite{Catani} Since the
carrier concentration can be controlled in experiments, we explore the doping dependence of the 
model to find complex and exotic phases~\cite{Elbio}.
 
While experimental controllability is a remarkable aspect of atomic systems,
the major goal of simulating quantum magnetism still remains a challenge. The 
main obstacle is reaching the low entropy and temperature required to observe magnetically 
ordered or Mott insulating phases. Various methods have been suggested in the last decade 
or so.~\cite{Monroe, Popp, Li}  A recent proposal by Ho and Zhou suggests that the entropy of 
a Fermi gas can be squeezed into a surrounding Bose-Einstein condensed gas, which acts as a 
heat reservoir.~\cite{Ho-Zhou}  These light particles are then evaporated, leaving behind a
low-entropy Fermi gas.  

In this Rapid Communication, we show that the two species bosonic Hubbard 
model with a mass imbalance at {\it finite} doping exhibits a ferromagnetic phase 
separated state, in addition to superfluid and antiferromagnetic phases.  This state 
has entropy similar to the superfluid indicating that it should have similar experimental 
accessibility.  Furthermore, it exhibits the entropy squeezing phenomenon 
in which the heavy particles form a Mott insulating phase, while the light ones 
are in a superfluid phase that act as a heat reservoir to absorb entropy. In addition, we study the  
entire phase diagram of this two species bosonic Hubbard model as a function of temperature and doping. 

Our study is based on the two-species Hubbard model with hard-core bosons {\it a} and {\it b} 
confined to a two-dimensional lattice.  The Hamiltonian takes the form:
\begin{eqnarray}
  \label{Hamiltonian} \nonumber \hat\mathcal H &=& -t_{a}\sum_{\big\langle i,
j\big\rangle}\Big(a_i^\dagger a_j^{\phantom\dagger}+H.c.\Big)\\
& & -t_{b}\sum_{\big\langle i,j\big\rangle}\Big(b_i^\dagger b_j^{\phantom\dagger}
+H.c.\Big)+{U^{ab}\sum_i  n_i^{a}  n_i^{b}},
\end{eqnarray}
where $a_i^\dagger$ ($b_i^\dagger$) and $a_i$ ($b_i$) are the creation and annihilation operators, respectively, of 
hard-core bosons {\it a} ({\it b}), with number operators $n_i^{a}=a_i^\dagger a_i^{\phantom\dagger}$, 
$n_i^{b}=b_i^\dagger b_i^{\phantom\dagger}$.  The sum $\sum_{\langle i,j\rangle}$ runs over all
distinct pairs of first neighboring sites $i$ and $j$, $t_a(t_b)$  is the hopping integral between $i$ and $j$ sites for 
species {\it a} ({\it b}), and $U^{ab}$ is the strength of the interspecies repulsion.   In the hard-core 
limit, the creation and annihilation operators satisfy commutation rules on different sites 
and anti-commutation
rules on identical ones.

We perform quantum Monte Carlo simulations using the Stochastic Green Function algorithm~\cite{SGF} with 
global space-time updates~\cite{SpaceTime} for the canonical ensemble on $L \times L$ lattices. We use an inverse temperature 
$\beta=8L$ to capture the ground state properties. Our results at half-filling reproduce the phase diagram 
of Ref.~\onlinecite{Soyler}.  We focus on the unpolarized phase diagram, so our total density is $\rho= N/L$ with 
$N=N_a+N_b= 2 N_a$, with $N_a$ and $N_b$ the number of heavy {\it a} and light {\it b} particles, respectively. 
We restrict our simulation to the following parameters corresponding to the strongly AF region at half filling: 
$t_a=0.08\, t$, $t_b=t$, and $U^{ab}=6t$, 
where $t=1$.
\begin{figure}[!htbp]
  \centerline{\includegraphics[width=0.5\textwidth]{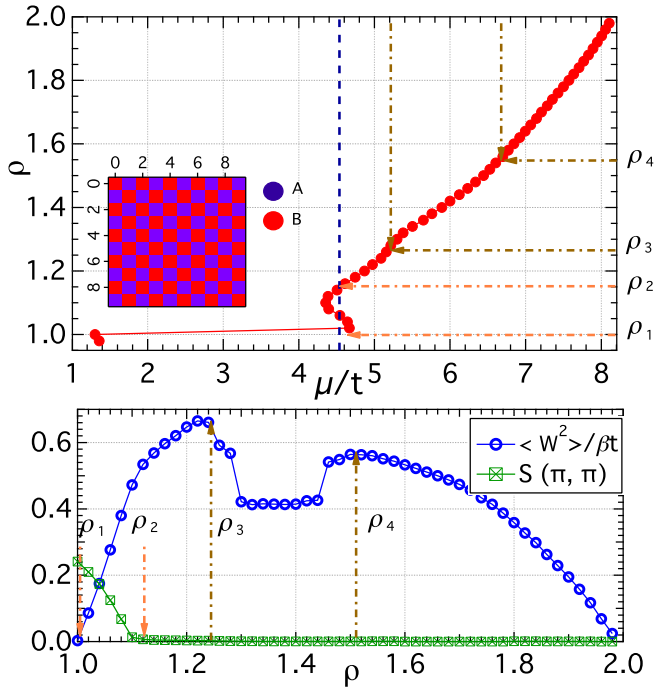}}
  \caption
    {(Color online) 
Top panel: The average density, $\rho=N/L$ as a function of the chemical potential, $\mu$. The vertical blue dashed line 
shows the Maxwell construction between $\rho_1$ and $\rho_2$. The brown dotted-dashed lines show different phase boundaries 
at $\rho_3$ and $\rho_4$ as discussed in the text. The inset shows a snapshot of the density profile at half filling,
$\rho=\rho_1=1$, with the blue (red) squares indicating $\langle n_i^{a} \rangle=1$ ($\langle n_i^{b} \rangle =1$).  Bottom panel:  The correlated winding
(blue circles) and the staggered structure factor (green squares) as a function of $\rho$. 
The dot-dashed lines show different phase boundaries. All data are for $L =10$, $\beta=80$, $t_a=0.08$, $t_b=1.00$ and 
$U^{ab}=6$.  Error bars are smaller than symbol sizes. 
   }
\label{Ground}
\end{figure}

Fig.~\ref{Ground} displays signatures of ordering.   To look for phase separation and Mott states, we calculate the 
chemical potential by adding one {\it a} and one {\it b} particle to the system as $\mu=(E(N+2)-E(N))/2$.  The superfluid
(SF) phase is detected by measuring the superfluid density, $\rho_{SF}$, using the fluctuations of the winding number, $W$, 
via Pollock and Ceperley's formula~\cite{Pollock}. We measure the correlated winding: $\langle W^2 \rangle=\langle (W_a+W_b)^2 \rangle$, where $W_a$ and $ W_b$ are the windings of particle {\it a} and {\it b} respectively. The AF phase is characterized by a finite density-density static structure factor:
$\displaystyle S(\vec k)=\frac{1}{L^{2}}\sum\limits_{k,l}exp[i\vec k \cdot (\vec r_k-\vec r_l)]\langle n_{k}^{(a,b)}n_{l}^{(a,b)}\rangle$.
We find an AF phase at half filling $\rho_1=1$.  It is characterized by a vanishing compressibility,
$\displaystyle \kappa=\partial\rho/\partial\mu$ (top panel), a finite static staggered structure factor (bottom panel),
as well as AF ordering as shown in a snapshot of the density profile (inset of the top panel). Near half filling, 
$\rho=N/L$ vs.\ $\mu$ displays a first-order phase transition between $\rho_1$ and $\rho_2\sim 1.16$. The instability 
is characterized by a region of negative slope. These two phases with densities $\rho_1$ and $\rho_2$ 
coexist for any density value between the two end points. Since, in this region the system displays finite values of  
$S(\pi, \pi)$ and correlated winding we conclude that the AF and SF phases coexist for any  $\rho_1 < \rho <\rho_2$. A homogeneous 
SF state exists between $\rho_2$ and $\rho_3 \sim 1.25$ identified by measuring the correlated winding $\langle W^2 \rangle$.
At $\rho_3$  the superfluid density displays a decrease and the $\rho$ versus $\mu$ plot shows a small bump.  Another 
small feature is displayed in the top panel at $\rho_4 \sim 1.52$.  Finally, the homogeneous SF phase continues until 
full filling. Next, we investigate the unexpected lowering of the superfluid density between $\rho_3$ and $\rho_4$.

\begin{figure}[!htbp]
  \centerline{\includegraphics[width=0.5\textwidth]{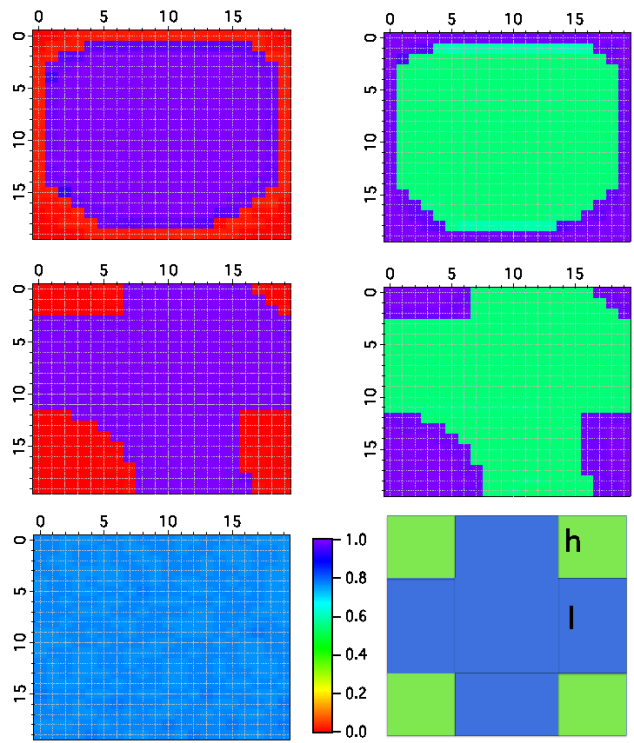}}
  \caption
    {
      (Color online) Snapshot of the average local densities for $L=20$ and $\rho=1.44$. 
Top panel: Open boundary conditions. 
For {\it a} particles (left panel), sites close to the boundary (red) have  $\langle n_i^a\rangle \sim 0$, while 
the occupation of the central (blue) region is $\langle n_i^a\rangle \sim 1$.
For {\it b} particles (right panel) the density close to the boundary (blue region) is $\langle n_i^b\rangle \sim 1$,  
and at the center (green region),  $\langle n_i^b\rangle \sim 0.60$.
Middle panel: The same quantities shown in the top panel but with periodic boundary conditions.  
Bottom panel: At the left the homogeneous density distribution of both  {\it a} and {\it b} particles for $\rho=1.72$.
At the right, a sketch of the density profile for a simulation with periodic boundary conditions.
   }
\label{Density}
\end{figure}

Fig.~\ref{Density} shows snapshots of the average local density of both species for $\rho_3<\rho=1.44<\rho_4$. 
Simulations with both open and periodic boundary conditions show clear evidence of ferromagnetic phase separation 
into regions with polarizations $\langle n^a_i-n_i^b\rangle$ of opposite sign.  The heavy species {\it a} shows Mott
behavior with integer fillings, $\langle n_i^a\rangle \sim 0$ or $1$, while the light species {\it b} shows Mott 
($\langle n_i^b\rangle \sim 1$) and SF ($\langle n_i^b\rangle \sim 0.60$) phases. We can understand the tendency 
of the system to form such a mixed phase by extending the bosonic mean-field formalism~\cite{Fisher, Sheshadri} to 
two species in the hardcore limit.~\cite{Altman} We use the Gutzwiller variational approach,~\cite{Rokhsar} in 
which the most general site factorized wave function can be written as
\begin{eqnarray}
  \label{Phi}  
\nonumber \Psi = \prod\limits_{i} \Psi_i &=&\prod\limits_{i}
\Big[\sin\frac{\theta}{2}(\sin\frac{\alpha}{2}a_i^\dagger+\cos\frac{\alpha}{2} b_i^\dagger)\\
& &+\cos\frac{\theta}{2}(\sin\frac{\beta}{2}+\cos\frac{\beta}{2} a_i^\dagger b_i^\dagger)\Big]|0\rangle,
\end{eqnarray}
where $\theta$, $\alpha$ and $\beta$ are variational parameters and $|0\rangle$ is the vacuum state.
The energy per site takes the form 
\begin{eqnarray}
  \label{energy}  
\nonumber \frac{E}{L^2} &=& 
-t_a \sin^2\theta \cos^2 \Big(\frac{\alpha-\beta}{2}\Big) -t_b \sin^2\theta \sin^2 \Big(\frac{\alpha+\beta}{2}\Big)\\ 
& & +U^{ab} \cos^2 \frac {\theta}{2}\cos^2\frac{\beta}{2}.
\end{eqnarray}
We solve these equations by minimizing the energy of the superfluid and phase separated 
states when $L=20$, $t_a=0.08$, $t_b=1.0$ and $\rho=1.44$, subject to the constraints
$\displaystyle n_a=\frac{N_a}{L^2}=\sum_i \frac{\langle \Psi | n_i^a | \Psi \rangle}{L^2}$ 
and $\displaystyle n_b=\frac{N_b}{L^2}=\sum_i \frac{\langle \Psi | n_i^b | \Psi \rangle}{L^2}$.
For $\rho=1.44$ we 
find that the phase separated state is lower in energy than a homogeneous superfluid phase.  
The total energy of a homogeneous superfluid with densities $\rho_a=\rho_b=0.72$ is $804.33$.   
The phase separated state is illustrated in the bottom right panel of Fig.~\ref{Density} with 
a central cross with area $(2h+l)^{2}-4h^{2}$ and four corner squares of area $h^{2}$. If we 
assume that in the cross $\langle n_i^a \rangle =1.0$ and $\langle n_i^b \rangle=0.60$ for all $i$ sites, then the mean-field 
energy is  $754.19$. 
The PS region is stabilized by the reduction of the potential energy, consequently there is a critical 
$U^{ab}_c$  above which the phase separated state is stable. We estimate $U^{ab}_c \sim 4.8$.

\begin{figure}[!htbp]
  \centerline{\includegraphics[width=0.5\textwidth]{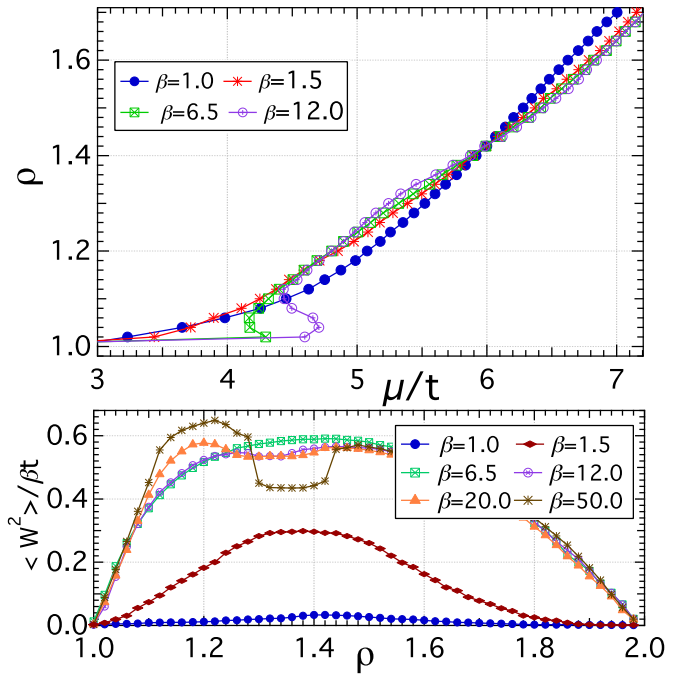}}
  \caption
    {
      (Color online)  
 Top panel:  Average density, $\rho$, versus chemical potential, $\mu$, for different temperatures.
 Bottom panel: The correlated winding as a function of $\rho$ for different temperatures. 
All data are for system size $L =10$. Error bars are smaller than symbol sizes.
    }
\label{Finite}
\end{figure}

Next we study the temperature dependence of  $\rho$ and $\rho_{SF}$ using the correlated winding.  
The top panel of Fig.~\ref{Finite} 
shows $\rho$ versus $\mu$ for a system of size $L=10$ and different inverse 
temperatures. Since there is a clear signature of phase separation for $\beta=6.5$ but not for $\beta=1.5$, we 
can estimate that the critical AF temperature occurs between these two temperatures. Similarly, from the 
$\langle W^2 \rangle$ vs.\ $\rho$ curves for different temperatures (see bottom panel of Fig.~\ref{Finite}), 
we can conclude that for this cluster size the phase separated region between $\rho_3$ and $\rho_4$ appears for 
temperatures between $\beta=6.5$ and $\beta=12$. We infer the phase diagram by appropriate scaling of our 
finite-size results.

\begin{figure}[!htbp]
  \centerline{\includegraphics[width=0.5\textwidth]{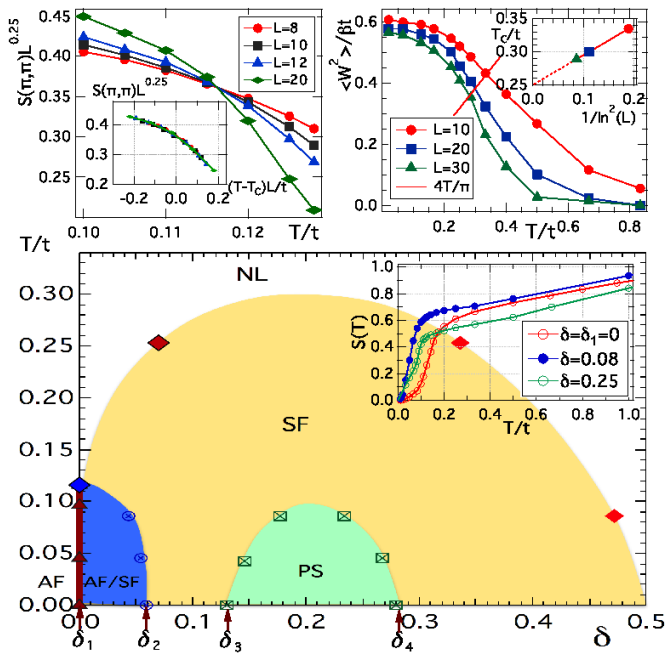}}
  \caption
    {(Color online)
Bottom panel: The temperature, $T$, versus doping, $\delta$, phase diagram with equal population of each 
species.  At half-filling, the system is antiferromagnetic (AF) below $T^{AF}_c \sim 0.116$ and normal 
liquid (NL) at larger temperatures. By increasing the doping a discontinuous transition from AF to superfluid 
(SF) phase occurs, and an AF/SF phase separated region develops between $\delta_1=0$ and $\delta_2=0.06$. 
Between $\delta_3=0.13$ and $\delta_4=0.28$ phase separation (PS) occurs with the heavy species becoming 
Mott while the light one displays regions with either Mott or superfluid behaviors. The inset shows entropy 
as a function of temperature for three dopings. Top left panel: Scaling behavior of the static staggered 
structure factor for the continuous transition from AF to NL at half filling, $\delta_1=0$ (corresponds to 
the filled blue diamond in the bottom panel).  The inset shows the scaling near the critical temperature 
with Ising-like critical exponents.  Top right panel: Correlated winding as a function of temperature for 
different system sizes at $\delta=0.07$ (filled red diamond). The inset shows the finite size scaling 
to find the SF critical temperature in the thermodynamic limit for the continuous transition at $\delta=0.07$. 
The data points are based on simulation results, the lines are guides to the eye.
    }
\label{Phasediagram}
\end{figure}

Fig.~\ref{Phasediagram} displays the temperature, $T$, vs.\ doping, $\delta=\displaystyle \frac{N_a+N_b}{2L^2}-\frac{1}{2}$, 
phase diagram. In the thermodynamic limit, the AF phase only exists at half filling $\delta_1=0$ and low temperatures. 
The top left panel shows the scaling of the AF to normal liquid (NL) continuous phase transition at  $\delta_1=0$.
This transition belongs to the two-dimensional Ising universality class for which the static staggered structure 
factor scales as $S(\pi, \pi)=L^{-(2\beta/\nu)}f((T-T^{AF}_c) L^{1/\nu})$, where $f$ is a universal scaling function, and 
$\beta$ and $\nu$ are the critical exponents for the order parameter and the correlation length, respectively. 
The factor $2\beta/\nu=1/4$ in two-dimensional systems. Therefore we can read the critical temperature at the point
where the  $S(\pi, \pi) L^{1/4}$ vs.\ $T$ curves for different system sizes cross.    
For our parameters  $T^{AF}_c=0.116$, and $S(\pi, \pi) L^{1/4}$ vs.\ $(T-T^{AF}_c)L$ curves collapse. 
The AF phase is represented by a red line ending on a blue diamond in 
Fig.~\ref{Phasediagram}.
As we illustrate in previous figures, near half filling, we find a discontinuous transition from AF to SF phases
and a phase separation region for doping $\delta_1=0.0 \lesssim \delta \lesssim \delta_2=0.06$ (dark blue region 
in  Fig.~\ref{Phasediagram}).
The boundary of the AF/SF phase separated region is found by a Maxwell's construction of the $\rho$ vs.\ $\mu$ plots. 
The PS region inside the SF phase exists for  $\delta_3=0.13 \lesssim  \delta \lesssim \delta_4=0.28$.
For a given temperature we determine its boundaries by estimating the filling where $\rho_{SF}$ starts decreasing 
($\rho_3$ in Fig.~\ref{Ground}) or stops increasing ($\rho_4$).
For the rest of the dopings we encounter a SF phase at low temperatures and an NL at higher temperatures.
The top right panel of  Fig.~\ref{Phasediagram}
shows the correlated winding as a function of temperature for different system sizes. 
The order parameter, the superfluid density, has the universal jump of $\langle W^2 \rangle=\displaystyle \frac{4}{\pi}$ 
at the critical point, $T_c(L)$.~\cite{Nelson} 
The transition from SF to NL belongs to the Kosterlitz Thouless universality class. 
We find $T_c$ in the thermodynamic limit by using the relation between the crossing temperature for different system sizes 
$T_c(L)$ and the cluster size: $T_c(L)-T_c(\infty)\propto \displaystyle \frac {1}{ln^{2}(L)}$.~\cite{Boninsegni05}
The inset on the right top panel  displays this scaling.
For $\delta=0.07$ ($\rho=1.14$) we find $T_c=0.25$. 
Scaled transition points are shown as red diamonds in Fig.~\ref{Phasediagram}. The inset of the bottom panel shows 
the entropy for a $L=10$ system calculated by following Ref.~\onlinecite{Werner} for $\delta=0$, $\delta=0.08$ and $\delta=0.25$.
The entropy of the PS ferromagnetic phase is greater than the AF phase and similar to the SF state, especially for 
low temperatures, indicating that it may be experimentally accessible. In addition, the entropy of this phase will mainly be carried by the light superfluid particles, enabling entropy squeezing.~\cite{Ho-Zhou}

In summary, with doping we find complex phases in the two-dimensional two-species hardcore bosonic Hubbard model 
for equal populations and unequal masses. We find a first order phase transition between the AF phase at half filling 
and a SF phase near half filling with a region of SF and AF coexistence. For a broad region of temperatures and fillings
away from half, a SF phase is found. Most significantly, within the SF phase at finite doping we find a dome-shaped 
region containing an inhomogeneous ferromagnetic phase. Density profiles of this novel phase separated region show that 
the heavy species displays Mott insulating behavior, while the light species are phase separated into Mott and superfluid 
regions.  Despite the magnetic order, this phase has an entropy much greater than the AF phase, and similar to the 
SF phase.  For a large system size, the entropy of the heavy species in this phase is essentially zero. This phenomenon 
can be considered as squeezing out the entropy from the heavy species into the light species, while both species are 
bosonic, in contrast with the recent proposal for cooling the boson-fermion mixture.~\cite{Ho-Zhou} Farther from half filling, 
a SF phase appears in a broad region of fillings at low temperature, while a NL phase appears at all fillings and high 
temperature. Further investigation of the phase diagram shows a rather complex SF phase which we discuss in  
future publications.

This complex phase diagram reminds us of the phase 
diagram of cuprates. In particular, our phase diagram displays a region that is similar to the so-called ``superconducting 
dome'' away from half-filling. Further, we believe our work will encourage experimental studies of this model on cold 
atoms traps. Indeed the experimental realization of the half-filling AF phase is difficult due to the low entropy 
associated with this phase, while the complex ferromagnetic phase that we identify away from half-filling can be expected 
to have a high entropy and, then, be easier to obtain. 
To further explore this complex phase diagram we are planning to extend our simulations to polarized 
systems with different population for each specie. 

We thank D.\ Browne and D.\ Galanakis for useful discussions. 
This work is supported by NSF OISE-0952300 (KH, VGR and JM).  Additional support was provided by DOE SciDAC grant 
DE-FC02-06ER25792 (KMT and MJ). 
This work used the Extreme Science and Engineering Discovery Environment (XSEDE), which is supported by the National 
Science Foundation grant number DMR100007, and the high performance computational resources provided by the Louisiana 
Optical Network Initiative (http://www.loni.org).

\end{document}